\pdfoutput=1
\documentclass[11pt,a4paper]{article}
\usepackage{amsmath,amssymb}
\usepackage{braket}
\usepackage{url}
\usepackage{mathtools}
\usepackage{amsmath}              
\usepackage{graphics,graphicx,epsfig,ulem} 
\usepackage{subfigure,subfig}
\usepackage{feynmp}
\usepackage{slashed}
\usepackage{xcolor}

\DeclareMathAlphabet{\mathbold}{U}{zeur}{b}{n}

\renewcommand\[{\left[}
\renewcommand\]{\right]}

\def\beq{\begin{equation}}
\def\eeq{\end{equation}}
\def\[{\begin{equation}}
\def\]{\end{equation}}

\newcommand{\startappendix}{
\setcounter{section}{0}
\renewcommand{\thesection}{\Alph{section}}
\renewcommand{\theequation}{\Alph{section}.\arabic{equation}}}

\newcommand{\Appendix}[1]{
\refstepcounter{section}
\begin{flushleft}
{\Large\bf Appendix: #1}
\end{flushleft}}

\begin{document}
\numberwithin{equation}{section}

\title{
{\normalsize  \mbox{}\hfill IPPP/16/41, DCPT/16/82}\\
\vspace{2.5cm}
\Large{\textbf{Multi-Higgs production in gluon fusion at 100 TeV}}}

\author{C{\'e}line Degrande, Valentin V. Khoze, Olivier Mattelaer\\[4ex]
  \small{\it Institute for Particle Physics Phenomenology, Department of Physics} \\
  \small{\it Durham University, Durham DH1 3LE, United Kingdom}\\[1ex]
  \small{\tt celine.degrande, valya.khoze, o.p.c.mattelaer@durham.ac.uk}\\[0.8ex]
}

\date{}
\maketitle

\begin{abstract}
 \noindent  
We carry out a detailed study of multi-Higgs production processes in the gluon fusion channel in the high energy regime relevant to Future Circular hadron colliders and in the high-Higgs-multiplicity limit ($\geq 20$). Our results are based on the computation of the leading polygons --the triangles, boxes, pentagons and hexagons-- to the scattering processes, further combined with the subsequent branchings to reach high final state multiplicities. The factorial growth of the number of diagrams leads to an exponential enhancement of such large multiplicity cross-sections and, ultimately, in breaking of perturbativity.
We find that the characteristic energy and multiplicity scales where these perturbative rates
become highly enhanced and grow with increasing energy are within the 100 TeV regime with of the order of 130 Higgses (or more) in the final state. We also show that already for a 50 TeV hadron collider the perturbative cross-sections for 140 bosons are at picobarn level.
 \end{abstract}

\bigskip
\thispagestyle{empty}
\setcounter{page}{0}

\newpage


\section{Introduction}\label{sec:intro}

At very high energies, production of multiple Higgs and electro-weak vector bosons becomes kinematically possible.
The cross-sections for such processes, computed in perturbation theory, become unsuppressed above a certain critical
value of final state multiplicities, and continue to grow with energy eventually violating perturbative unitarity
\cite{Cornwall,Goldberg,Voloshin80}.
This results in the breakdown of a weakly
coupled perturbation theory and a transition into a non-perturbative regime where the intermediate state 
formed during the collision is characterised by a collective
multi-boson configuration with large occupation numbers.
The critical energies were the perturbative high-multiplicity rates become large
were estimated recently in \cite{Jaeckel:2014lya,VVK3} and were found to be in the
$10^2$ - $10^3$ TeV range -- i.e. nearly in the reach of current and future experiments.
The aim of this paper is to further improve and quantify the critical values of energies and multiplicities for the multi-Higgs boson production.
We will argue that already for a 50 TeV hadron collider, the rapid growth of perturbative rates in our 
model can lead to picobarn cross-sections for processes with $\gtrsim 140$ Higgs bosons.

\medskip

There is a strong similarity, already noted in  \cite{Jaeckel:2014lya}, between these novel perturbative unitarity problems 
at high multiplicities with hundred TeV energies, and the well-known unitarity problem 
for simple 2 to 2 scattering processes of massive vector bosons. This resulted in a powerful and far-reaching  
conclusion formulated in \cite{Lee:1977yc} that 
one of the three options has to be realised:  either (1) there 
exists a Higgs boson with a mass below $\sim1$~TeV, or (2) there should be new physics beyond the Standard Model, or, finally (3)
scattering processes of electroweak gauge bosons become non-perturbative. This three-fold way forward for electro-weak
physics was answered and resolved by the observation of a Higgs boson at $125$~GeV. Now, with the high-multiplicity
scatterings at $50-100$ TeV centre of mass energies, the perturbative electro-weak physics faces a similar cross-roads.

\medskip

To obtain a reliable estimate for multi-Higgs production processes 
at energies relevant for future circular hadron colliders (FCC) which kinematically allow for very high Higgs multiplicities in the final state,
one has to overcome a number of complications. 
There are two immediate technical problems one encounters already at the leading
order in perturbation theory:

\begin{enumerate}
\item The dominant Higgs production is via the gluon fusion process $gg \to n\times h$, and it requires a computation 
of Feynman diagrams involving 
1-loop polygons with $2+k$ edges where $k$ is the number of the outgoing Higgs lines, for all $k\le n$. 
The number of the contributing polygon types and of the corresponding kinematic invariants they depend on, grow with $n$
and ultimately explode in the high multiplicity limit $n\gg1$.
This provides for a challenging computation. 
\item The number of Feynman diagrams describing the subsequent tree-level 
branching processes $h_i^* \to n_i\times h$ from each of the polygon's external lines $h_i^*$, is known to grow factorially with $n$, and this is reflected in a factorial explosion of perturbative amplitudes, as shown in 
Refs.~\cite{Brown,Argyres73,LRST,VVK1}.

Based on these considerations it was argued in \cite{Jaeckel:2014lya,VVK3}
that the standard weakly coupled perturbation theory in the electro-weak
sector of the Standard Model breaks down for multi-particle production of Higgses and massive vector bosons at energy scales as low
as $\sim 10^2-10^3$ TeV. The energies where electro-weak processes could enter a novel effectively strongly-coupled regime, where 
the ultra-high multiplicity production of relatively soft bosons would become unsuppressed and dominate the total rates may be potentially within the reach of the next generation of colliders.
\end{enumerate}

We will address the two problems listed above in stages: first we will consider the polygon contributions to the multi-Higgs cross-sections
by working in the high-energy limit $\sqrt{s} \to \infty$ with a fixed number of Higgses, $k=$fixed. Then we will combine these
fixed-multiplicity loop-level results in the  ultra-high-energy limit with the subsequent tree-level branchings. Here each intermediate highly energetic 
Higgs particle $h^*_i$ emitted at the end of the polygon-production stage, undergoes the tree-level production  $h^*_i\to n_i\times h$
into the high multiplicity $n$-Higgs final state, $n=\sum_i n_i$. The full amplitude chain for this process is,
\[
{\cal A}_{gg \to n\times h}\,=\, \sum_{\rm polygons} {\cal A}^{\rm polygons}_{gg \to k\times h^*}\,\,
 \sum_{n_1+\ldots+n_k=n}\, \prod_{i=1}^k \,{\cal A}_{h_i^* \to n_i\times h}\,.
\label{eq:chain}
\]
The $1_i^*\to n_i$ amplitudes\footnote{We will always adopt the 
short-hand convention that the propagator for the incoming virtual Higgs was not LSZ amputated, i.e.
$ {\cal A}_{h_i^* \to n_i\times h} := \frac{1}{s_i-M_h^2}\, {\cal A}_{h_i^* \to n_i\times h}^{LHZ}$.}
 appearing as the right-most factor in \eqref{eq:chain}
can be computed very efficiently for all $n_i$ using the classical generating functions technique. For convenience
and future reference in we will now present the result for these amplitudes on multi-Higgs mass thresholds. 

The computation of polygons contributions to the processes \eqref{eq:chain} combined with the subsequent branchings and the 
resulting estimate
for the multi-Higgs production cross-sections, which is the main motivation of this paper, will be addressed
in Sections {\bf \ref{sec:poly}}-{\bf \ref{sec:PDFs}}.

\subsection{ ${\cal A}_{h^*\to n\times h}$ from classical solutions}\label{sec:I1}

At tree-level, all $n$-point scattering amplitudes for an off-shell field $h$ to produce $n$ Higgs particles, ${\cal A}_{1\to n}$,
can be obtained from a classical solution of the Euler-Lagrange equations corresponding to 
the Higgs Lagrangian
\[
{\cal L}_h \,= \, \frac{1}{2}\, \partial^\mu h \, \partial_\mu h\, -\,  \frac{\lambda}{4} \left( h^2 - v^2\right)^2
\,,
\label{eq:LSSB}
\]
following the generating functions technique initiated in Ref.~\cite{Brown} ($\lambda$ is the Higgs self-coupling and $v$ the vacuum expectation value). For an overview of the classical generating functions technique and its applications, the interested Reader can consult the Appendix. In the rest of the current section we will simply state the features of this approach which are relevant for our study.

As the final state is made out of the outgoing particles, the relevant solution $h_{\rm cl} (x)$ should contain 
only the positive frequency modes, $e^{+in M_h t}$ where $M_h = \sqrt{2\lambda}\,v$ is the Higgs boson mass.  
This specifies the initial conditions, or equivalently the analytic structure of
the solution -- its time-dependence is described by the complex variable $z$, 
\[
z(t)\,=\,z_0 \, e^{iM_h t}\,,
\]
on which the configuration $h_{\rm cl}$
depends holomorphically,
\[
h_{\rm cl} (\vec{x},t) \,=\, v\,+\, \sum_{n=1}^{\infty} a_n(\vec{x})\, z(t)^n
\, .
\label{sol-T}
\]
As there is no dependence on the complex conjugate variable $z^*$, the required solution is complex (even though the original scalar filed 
$h$ was real) and will also contain singularities in the Euclidean space-time.

We now consider the simplest kinematics, where all the final state particles are produced at their mass threshold 
(i.e. with vanishing spacial momenta). In this case,
the classical solution in question, $h_{\rm cl}$, is uniform in space and solves the ordinary differential equation,
\[
d_t^2 h \,=\, -\lambda\,h^3 +\lambda v^2\,h
\,,
\label{cleq-SSB}
\]
with the initial conditions,  $h_{\rm cl} = v + z + {\cal O}(z^2)$. This solution is known in closed-form
\cite{Brown}:
\[
h_{\rm cl} (t) \,=\, v\,\frac{1+\frac{z(t)}{2v}}{1-\frac{z(t)}{2v}} \,.
\label{sol-SSB}
\]
This is the exact solution of the classical equation \eqref{cleq-SSB}, as can be readily checked for example in Mathematica.
In the Appendix we also explain how to derive this expression analytically. 

We note that (in real time) this expression is complex and that it is singular on the complex time plane at $z=2v$.
The singularity of the solution 
is the consequence of the finite radius of convergence of the Taylor expansion of \eqref{sol-SSB},
\[
h_{\rm cl} (t) \,=\,  v\,+\, 2v\,\sum_{n=1}^{\infty} \left(\frac{z(t)}{2v}\right)^n
\, .
\label{gen-funh}
\]
\medskip
The classical solution
$h_{\rm cl}$ defines the generating functional for the tree-level scattering amplitudes.
All $n$-point tree-level amplitudes at threshold are simply given by differentiating $n$-times with respect to $z$,
\cite{Brown}
\[
{\cal A}_{1\to n}\,=\, 
\left.\left(\frac{\partial}{\partial z}\right)^n h_{\rm cl} \,\right|_{z=0}
\,=\, n!\, (2v)^{1-n}
\,,
\label{eq:amplnh}
\]
and exhibit the factorial growth with the number of particles in the final state. 
Equations \eqref{sol-SSB} and \eqref{eq:amplnh} describing the tree-level amplitudes on the multi-particle mass thresholds,
will play an important role in our approach in the main part of this paper.\footnote{These are exact results for
the tree-level $n$-point amplitudes on mass thresholds for arbitrary values of $n$.}

Such threshold amplitudes were further generalised to other scalar field theories and also 
computed at one-loop and resummed multi-loop level in 
Refs.~\cite{Argyres:1992nqz,Voloshin83,Smith84,Argyres:1992kt,Makeenko:1994pw,Libanov:1996vq}.
The multi-particle amplitudes on threshold were also computed in the gauge-Higgs theory in \cite{VVK1}, confirming their
factorial growth in reference to probing the electroweak sector at high FCC energies \cite{Jaeckel:2014lya,VVK3}.

\medskip

For general kinematics, with momenta above the multi-particle mass threshold, the scattering amplitudes ${\cal A}_{1\to n}$ 
at tree-level are still given by the classical
solution $h_{\rm cl} (\vec{x},t)$ of equations of motion -- they are no longer uniform in space, having instead the $O(3)$ spherical symmetry.
These solutions are uniquely specified by the same initial conditions as $z\to 0$,  and are
singular on hypersurfaces in the Euclidean space-time. 
They could be found numerically by searching for classical extrema of the path integral on the appropriate
singular complex-valued field configurations as explained in  \cite{Son,Bezrukov:1995qh,Bezrukov:1999kb,LRT}.
This is a complicated procedure, and the
closed-form expressions for such $O(3)$ symmetric solutions are presently unknown
even in the simplest scalar QFT models.

Alternatively, one can derive the amplitudes and cross-sections dependence on the external states kinematics at tree-level
by solving the full $(3+1)$-dimensional Euler-Lagrange equations recursively in $n$.
This is achieved by writing down the perturbative recursion relations corresponding to the classical solutions, as explained in
Refs.~\cite{LRST,LST,VVK2}, and solving them 
first in the non-relativistic limit, and then in general kinematics. The latter step is required to enable the integration over the  
$n$-particle phase space to obtain the cross-section. 
This programme was carried out in Ref.~\cite{VVK3} 
using MadGraph5\_aMC@NLO \cite{Alwall:2011uj,Alwall:2014hca}.
The approach followed in this paper will not require the knowledge of the $\vec{x}$-dependent singular solutions, instead we will 
use the formalism and results of \cite{VVK3}
based on combining the known scaling behaviour at large $n$ inferred from the mass-threshold amplitude \eqref{eq:amplnh},
with a numerical computation of tree-level cross-sections at fixed $n$ directly.

\medskip

This paper is organised as follows.
In Section {\bf \ref{sec:poly}} we will compute the gluon fusion cross-sections for the double, triple, and quadruple Higgs production 
at fixed center-of-mass gluon energies in the range between 10 and 160 TeV. We will identify the contributions coming from 
the triangles, boxes, pentagons and hexagons and represent them in the high-energy regime in terms of effective
vertices with energy-dependent form-factors. We will demonstrate that this approximation is well-justified in the high-energy
kinematics where $\sqrt{s}$ is much greater than masses of the Higgs and the top quark. We  will then combine the
effective vertices with the classical generating functions for tree-level amplitudes describing the subsequent
multi-Higgs branchings. In this way we will obtain the generating functions for scattering amplitudes describing
$gg \to n\times h$ processes in the high multiplicity regime near the multi-particle mass thresholds.
We will use these results in Section~{\bf \ref{sec:An}} to estimate the multi-particle cross-sections 
based on their scaling behaviour with multiplicity and energy \cite{LRST,VVK3}. 
Finally in Section~{\bf \ref{sec:PDFs}} we will convolute the partonic cross-sections with the parton distribution functions (PDFs) of the 
gluons. Our projections for the high-multiplicity Higgs production cross-sections at proton-proton colliders 
are summarised in Fig.~\ref{fig:pdf} and our conclusions are presented in Section~{\bf \ref{sec:conc}}.

\medskip

\section{Polygons and effective vertices in the $\sqrt{s}\to \infty$ limit}\label{sec:poly}

 \begin{figure*}[t]
\begin{center}
\begin{tabular}{c}
\includegraphics[width=0.48\textwidth]{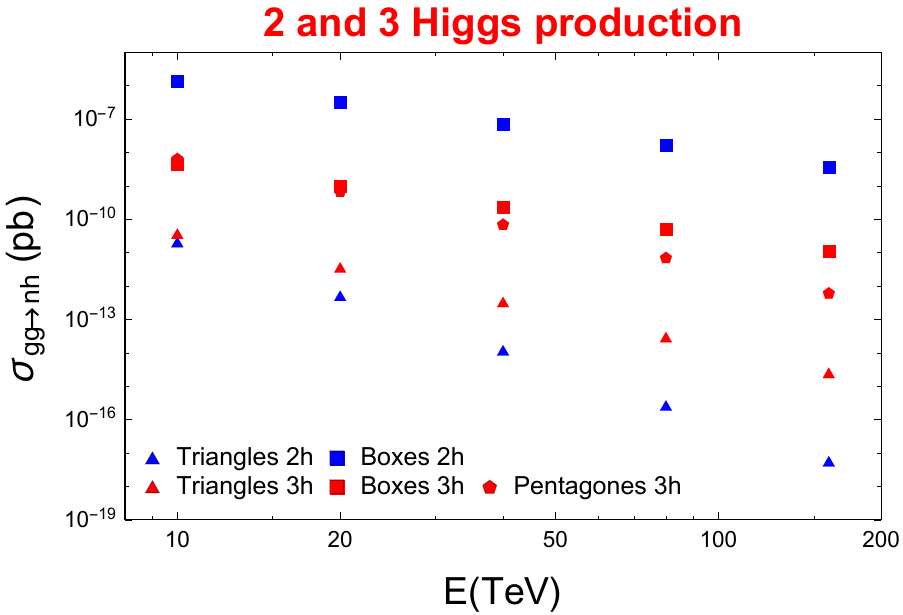}
\includegraphics[width=0.48\textwidth]{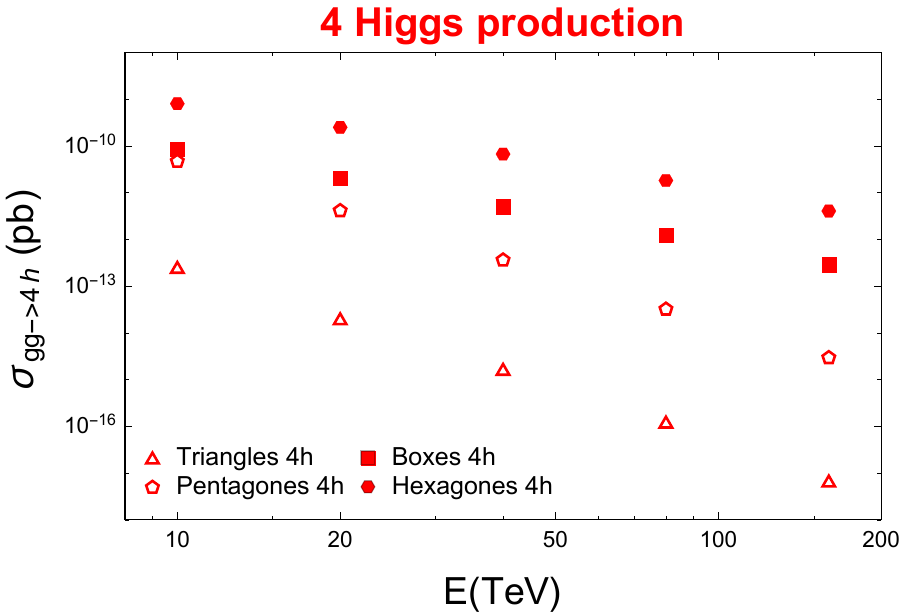}
\\
\end{tabular}
\end{center}
\vskip-.6cm
\caption{
Cross-sections for the 2-Higgs, 3-Higgs and 4-Higgs production in the gluon fusion process separated into contributions 
from triangles, boxes, pentagons and hexagons, as indicated. Gluons are scattered at fixed energy (i.e. no gluon PDFs included) 
in order to simplify the $s$-dependence
of these cross-sections at partonic level.
}
\label{fig:poly}
\end{figure*}

We now consider the first stage of the process \eqref{eq:chain}
involving the high-energy fixed multiplicity $k$-Higgs production $ {\cal A}^{\rm polygons}_{gg \to k\times h^*}$.
The double and triple Higgs production at colliders was studied in 
Refs.\cite{Glover:1987nx,Plehn:1996wb,Binoth:2006ym,Borowka:2016ehy} 
and \cite{Papaefstathiou:2015paa,Fuks:2015hna,Baglio:2015wcg,Chen:2015gva}
and is rather suppressed at the LHC and the FCC energies. Our main goal, however, is to determine whether the 
{\it high multiplicity} rates with $n\gg 2,3$ Higgses can become unsuppressed in perturbation theory. As explained in the Introduction, 
we will address the large-$n$ limit by computing the fixed multiplicity  ${gg \to k\times h^*}$ 1-loop processes 
in the high-energy limit and combine them with the subsequent $h^* \to n_i \times h$ branchings, {\it cf.} Eq.~\eqref{eq:chain}.

\medskip

Using the 
MadGraph5{\_}aMC{@}NLO framework \cite{Hirschi:2015iia}
we computed the double, triple and quadruple Higgs production cross-sections in the gluon fusion channel at 1-loop level
in the high-energy regime. Specifically, with the applications to the FCC hadronic colliders in mind, 
we concentrate on the centre of mass energies $\sqrt{s}$ much greater than the Higgs and top quark masses.

\begin{table}[t]
\begin{center}
\begin{tabular}{|c||c|c|c|}
\hline
$\quad$ & $\sigma_{gg\to\, hh}$ & $\sigma_{gg\to\, hhh}$  & $\sigma_{gg\to\, hhhh}$ \\
\hline \hline
&&&\\
Triangles & $ y_t^2\frac{m_t^2 M_h^2 }{s^3}  \log^4\left(\frac{m_t}{\sqrt{s}}\right) \frac{M_h^2}{v^2}$ 
& $  y_t^2\frac{m_t^2}{s^2}  \log^4\left(\frac{m_t}{\sqrt{s}}\right) \frac{M_h^4 }{v^4}$ 
& $y_t^2\frac{m_t^2}{s^2}  \log^4\left(\frac{m_t}{\sqrt{s}}\right) \frac{M_h^6 }{v^6}$ \\
&&&\\
Boxes &$y_t^4\frac{1}{s}$ & $y_t^4\frac{1}{s}\, \frac{M_h^2}{v^2}$  &$y_t^4\frac{1}{s}\, \frac{M_h^4}{v^4}$ \\
&&&\\
Pentagons & -- & $  y_t^6\frac{m_t^2}{s^2}  \log^4\left(\frac{m_t}{\sqrt{s}}\right) $ 
& $  y_t^6\frac{m_t^2}{s^2}  \log^4\left(\frac{m_t}{\sqrt{s}}\right) \frac{M_h^2}{v^2}$ \\
&&&\\
Hexagons & -- & --  &$y_t^8\frac{1}{s}$ \\
&&&\\\hline
\end{tabular}
\end{center}
\caption{High-energy scaling behaviour of each polygon-type contributions to the gluon fusion multi-Higgs production cross-sections,
extracted from numerical data as in Fig.~\ref{fig:poly}, and shown
as the function of $s$, $M_h$ and $m_t$ and $y_t:= \sqrt{2}m_t/v$ in the $s\gg m_t\,,\,  M_h$ limit. All cross-sections also contain the common factor of $\alpha^2_s(\sqrt{s})$. Gluon PDFs are not included.}
\label{tab:poly}
\end{table}

\medskip

The first panel in Figure~\ref{fig:poly} shows our results for the Higgs pair production and the triple Higgs production,
and the second panel gives the cross-sections for the quadruple Higgs.  
The contributions from each 
type of polygons are shown separately (and we do not compute the interference terms between different polygon types). For example, the triangles category corresponds to the sum of all Feynman diagrams 
containing the $gg \to h^*$ 1-loop triangles  contributing
to the $gg \to h^*\to n\times h$ amplitude for $n=2,3,4$. 
The resulting amplitude is squared and integrated over the phase space to obtain the
cross-section contributions induced by the triangles. The process is then repeated for higher polygons: boxes, pentagons and 
hexagons.\footnote{To be clear, in our notation the polygon ranks (i.e. the number of polygon edges) is $2+k$ where 
$1\le k \le n$ so that e.g. pentagons ($k=3$) contribute to $gg \to 3\times h^*\to n\times h$ processes with $n=3,4,\ldots$}

The interference terms between polygons with different numbers of sides (e.g. interferences between the triangles-induced and the boxes-induced contributions to the cross-sections) are not
accounted in the computation presented in Fig.~\ref{fig:poly}. 
However, based on the fact that different polygon types give a very clear 
numerical hierarchy of the cross-sections values, as seen from Fig.~\ref{fig:poly}, (and similarly have the 
different analytic dependence on the parameters,
as will be seen in Tables 1 and 2 below) we expect that the missing interference terms will not modify our results dramatically.

Varying the Higgs and top masses as well as the centre of mass energy $\sqrt{s}$ we can extract from these data the analytic scaling
properties for different polygonal contributions to the cross-sections applicable in the high-energy regime. These scaling properties are summarised
in Table~\ref{tab:poly}.
The polygons with different numbers of edges
are treated separately, so that the different entries in the Table do not mix e.g. triangles with boxes; each horizontal entry is 
specific to a particular type of polygons as indicated and contains no cross-terms between polygons with different numbers of edges.
We have also fixed the energy of the gluon (i.e. we are considering partonic cross-sections with no gluon PDFs)
in order to focus on the $s$-dependence
of the cross-sections at partonic level.\footnote{The proton-proton collisions and the convolution of the partonic
cross-sections with gluon PDFs will be discussed in Section {\bf \ref{sec:PDFs}}.}
It follows that all even polygons (boxes, hexagons, etc) exhibit the same $1/s$ scaling in the high-energy limit $\sqrt{s} \gg M_h\,,\, m_t$.
At the same time the odd polygons (triangles, pentagons, and so on) are sub-dominant and go as $1/s^2 \log^4(m_t/\sqrt{s})$ 
(with the exception of the leading double-Higgs case where the suppression is even stronger).

The high-energy behaviour of the leading-rank polygons in Table~\ref{tab:poly} can now be easily generalised to higher multiplicities
and higher polygon ranks following the same pattern. For polygons with $2+k$ edges their contribution to the 
$gg \to n\times h$ process is:
\[
(2+k){\rm -polygons:} \qquad
\sigma_{gg \to n\times h} \propto \, 
\frac{1}{s} \, y_t^{2k} \left(\frac{M_h}{v}\right)^{2(n-k)}
\times
\begin{cases}
1 & :\,\,k={\rm even}\\
\frac{m_t^2}{s} \log^4\left(\frac{m_t}{\sqrt{s}}\right) & :\,\,k={\rm odd}\, .
\end{cases}
\label{eq:poly_ev_odd}
\]
The only exception from this rule is the $k=1$, $n=2$ case, i.e. left-most triangle in Table~\ref{tab:poly},
which has an additional factor of $M_h^2/s$. As a matter of fact, the squared amplitude in multi-Higgs production with a odd number of three Higgs vertices is enhanced compared to a naive counting by a factor $s/M_h^2$ when the invariant mass appearing in the propagator is close to its minimal value of order $M_h$. In the case of pair production, the only invariant mass is fixed at $\sqrt{s}$ and therefore such enhancement is absent.

\medskip

The pattern established in Table~\ref{tab:poly}  and Eq.~\eqref{eq:poly_ev_odd} 
enables us to simplify the full 1-loop Feynman diagrams-based computation in  Fig.~\ref{fig:poly}    
by reducing it to contributions from effective 
multi-Higgs vertices of the form
\[
V^{\rm eft}_k \sim \alpha_s {\rm tr}(G_{\mu\nu}G^{\mu\nu}) \, h^k\,,
\label{eq:eft}
\]
where $\sim$ indicates that the dimension-$(4+k)$ operators on the right hand side
should be multiplied by the appropriate energy-dependent form-factors $F_k(s)$. These form-factors will be determined momentarily. 

To proceed we first consider the contributions to cross-sections from the bare effective operators \eqref{eq:eft}, i.e.
not including the form-factors. The corresponding cross-sections are found to grow with $s$, as summarised in Table~\ref{tab:eft},
and this is of course also consistent with a simple dimensional analysis in the high-energy limit.
\begin{table}
\begin{center}
\begin{tabular}{|c||c|c|c|}
\hline
$\quad$ & $\sigma^{\rm eft}_{gg\to\, hh}$ & $\sigma^{\rm eft}_{gg\to\, hhh}$  & $\sigma^{\rm eft}_{gg\to\, hhhh}$ \\
\hline \hline
&&&\\
 $\alpha_s {\rm tr}(G_{\mu\nu}G^{\mu\nu}) \, h^1$ & $ \frac{M_h^2}{v^2}s^0$ 
& $ \frac{M_h^4}{v^4}s^0$ 
& $\frac{M_h^6}{v^6}s^0$ \\
&&&\\
 $\alpha_s {\rm tr}(G_{\mu\nu}G^{\mu\nu}) \, h^2$ & $s$ 
& $ \frac{M_h^2}{v^2} s$ 
& $\frac{M_h^4}{v^4}s$ \\
&&&\\
 $\alpha_s {\rm tr}(G_{\mu\nu}G^{\mu\nu}) \, h^3$ &--& $s^2$  &$\frac{M_h^2}{v^2}s^2$ \\
&&&\\
 $\alpha_s {\rm tr}(G_{\mu\nu}G^{\mu\nu}) \, h^4$ & -- & -- & $s^3$  \\
&&&\\\hline
\end{tabular}
\end{center}
\caption{High-energy scaling behaviour for multi-Higgs production cross-sections with the bare effective vertices  Eq.~\eqref{eq:eft} obtained with FeynRules~\cite{Alloul:2013bka} and Madgraph5\_aMC@NLO. 
}
\label{tab:eft}
\end{table}
The form-factors $F_k(\sqrt{s})$ can now be determined by matching the contributions from ${\cal V}_k:= V^{\rm eft}_k \times F_k(\sqrt{s})$ of  Table~\ref{tab:eft}
to Table~\ref{tab:poly}. 
We find the following expressions for the effective vertices [including the form-factors]:
\[
{\cal V}_k =\, C_k \,\frac{\alpha_s(\sqrt{s})}{\pi}\, {\rm tr}(G_{\mu\nu}G^{\mu\nu}) \,\left(\frac{ y_t\, h}{\sqrt{s}}\right)^k
\times
\begin{cases}
1 & :\,\,k={\rm even}\ge 2\\
\frac{m_t}{\sqrt{s}} \, \log^2\left(\frac{m_t}{\sqrt{s}}\right) & :\,\,k={\rm odd}\ge 3\, .
\end{cases}
\label{eq:efteven}
\]
%
Here $C_k$'s are the constant coefficients to be determined by matching to the full numerical cross-section results, and $y_t$ is the top quark Yukawa coupling.
 
The coefficients  $C_k$ can now be found from matching the cross-sections $\sigma^{\rm eft}$ computed from the Effective Field Theory
(EFT) vertices \eqref{eq:efteven} to our numerical results for the complete partonic cross-sections shown in Fig.~\ref{fig:poly}.
Specifically the two-point effective vertices are matched to boxes, 
the tree-point EFTs -- to pentagons, and the four-point vertices are matched to 
the hexagon-induced contributions to the cross-sections. For each effective vertex of rank $k$, the coefficient $C_{k}$ can be obtained 
in $n-k$ independent ways
from matching:
\begin{eqnarray}
C_2: \qquad  \sigma_{gg\to\, n\times h}\,[{\cal V}_2] &\longleftrightarrow & \sigma_{gg\to\, n\times h}\,[{\rm Boxes}]\,,\,\, \,\,\,\qquad {\rm for} \,\, n=2,3,4,\ldots
\\
C_3: \qquad  \sigma_{gg\to\, n\times h}\,[{\cal V}_3] &\longleftrightarrow & \sigma_{gg\to\, n\times h}\,[{\rm Pentagons}]\,, \quad {\rm for} \,\, n=3,4,\ldots
\\
C_4: \qquad  \sigma_{gg\to\, n\times h}\,[{\cal V}_4] &\longleftrightarrow & \sigma_{gg\to\, n\times h}\,[{\rm Hexagons}]\,, \,\,\quad {\rm for} \,\, n=4,\ldots
\end{eqnarray}
and for different values of the centre of mass energy $\sqrt{s}$. Their values are shown in Table~\ref{tab:Cn}.
\begin{table}[h!]
\begin{center}
\begin{tabular}{|c||c|c|c|c|c|}
\hline
$C_2$ & $\sqrt{s}=10$ TeV & $\sqrt{s}=20$ TeV & $\sqrt{s}=40$ TeV & $\sqrt{s}=80$ TeV & $\sqrt{s}=160$ TeV\\
\hline \hline
$gg\to\, hh$ & 1.12 & 1.13 & 1.14 & 1.14 & 1.14\\\hline
$gg\to\, hhh$ & 1.11 & 1.13 & 1.14 & 1.14 & 1.14\\\hline
$gg\to\, hhhh$ &  1.21 & 1.23& 1.23 &1.23 & 1.21 \\\hline
\end{tabular}
\vskip0.6truecm
\begin{tabular}{|c||c|c|c|c|c|}
\hline
$C_3$ & $\sqrt{s}=10$ TeV & $\sqrt{s}=20$ TeV & $\sqrt{s}=40$ TeV & $\sqrt{s}=80$ TeV & $\sqrt{s}=160$ TeV\\
\hline \hline
$gg\to\, hhh$ & 7.91 & 7.95 & 8.16 & 8.59 &8.60\\\hline
$gg\to\, hhhh$ &  8.52 & 8.42 & 8.43 & 8.68 & 8.86 \\\hline
\end{tabular}
\vskip0.6truecm
\begin{tabular}{|c||c|c|c|c|c|}
\hline
$C_4$ & $\sqrt{s}=10$ TeV & $\sqrt{s}=20$ TeV & $\sqrt{s}=40$ TeV & $\sqrt{s}=80$ TeV & $\sqrt{s}=160$ TeV\\
\hline \hline
$gg\to\, hhhh$ &  4.34 & 5.10 & 5.55 & 6.11 & 6.04 \\\hline
\end{tabular}
\end{center}
\caption{Operator coefficients in Eq.~\eqref{eq:efteven}. Each $C_k$ appears to be largely independent 
of the number of Higgses in the full matching process $gg\to\, n\times h$ and describes the rates well at all energies in
the high-energy range $\sqrt{s}\gg M_h, m_t$.
}
\label{tab:Cn}
\end{table}
We conclude that the extracted numerical values of these coefficients do not appear to depend strongly on the number of 
Higgses in the final state.
This is an important test for our approach; it guarantees the robustness of the effective vertices approximation 
\eqref{eq:efteven} for the multi-Higgs production cross-sections in the high energy limit.

\medskip

Our construction up to this point was derived from taking the high-energy limit and holding the Higgs multiplicity fixed. 
The next step is to use of the effective vertices \eqref{eq:efteven} combined with the classical 
generating functionals for the tree-level amplitudes introduced in Section {\bf \ref{sec:I1}},
to address the desired high multiplicity limit $n \gg 1$.
This is achieved by substituting the Higgs fields $h$ in the effective vertices by the generating functionals for $1^*\to n$ 
scattering amplitudes ${\cal A}_{h^*\to n\times h}$.

The resulting generating functionals for the two gluons into any number of Higgses processes are given 
by the effective vertices ${\cal V}_k$ in Eqs.~\eqref{eq:efteven} with the substitutions $h^k \to h_{\rm cl}[z]^k$ and 
${\rm tr}(G_{\mu\nu}G^{\mu\nu}) \to (p_{1\mu}\epsilon_{1\nu}^a-p_{1\nu}\epsilon_{1\mu}^a)
(p_{2}^{\mu}\epsilon_{2}^{a\nu}-p_{2}^{\nu}\epsilon_{2}^{a\mu})$.
Here $p_1$, $p_2$ are the gluon momenta and $\epsilon_1^a$, $\epsilon_2^b$ are their helicities, while 
$h_{\rm cl}[z]$ is the generating functional \eqref{sol-T}.

Using the classical solution \eqref{sol-SSB} 
(shifted by the vacuum expectation value)
 $h_{\rm cl}\to h_{\rm cl} -v$,
we can immediately write down the 
the generating functional of multi-Higgs amplitudes on the multi-particle mass threshold in closed form:
\[
{\cal A}^{k={\rm even}} [z] =\, C_k \,\frac{\alpha_s}{\pi}\,(p_{1\mu}\epsilon_{1\nu}^a-p_{1\nu}\epsilon_{1\mu}^a)
(p_{2}^{\mu}\epsilon_{2}^{a\nu}-p_{2}^{\nu}\epsilon_{2}^{a\mu})\,\left(\frac{ y_t}{\sqrt{s}}\,\frac{z}{1-\frac{z}{2v}} \right)^k\,.
\label{eq:Aeft}
\]
Here we took the polygons/EFT vertex rank $k$ to be even-valued, since 
for odd $k$ the effective vertices in \eqref{eq:efteven} are suppressed by the factor
 $\frac{m_t}{\sqrt{s}} \, \log^2\left(\frac{m_t}{\sqrt{s}}\right) \ll 1$.

The high-multiplicity regime of interest for us is 
\[
\sqrt{s} \gg {\rm all\,\,other\,\,mass\,\,scales}\,,\quad {\rm and} \,\,\, n\gg1\,,
\]
and it will also be convenient to define the average final state kinetic energy  per particle per mass, $\varepsilon$, via 
\[
\varepsilon \,:=\, \frac{\sqrt{s}-nM_h}{nM_h} \,.\label{eq:eps_n}
\]
On the multi-particle mass-threshold $\varepsilon =0$, but more generally, above the threshold we will work in the limit
where $\varepsilon$ is held fixed at some non-vanishing value\footnote{Corresponding to either a non-relativistic ($\varepsilon < 1$)
or a highly relativistic ($\varepsilon > 1$) regime.}
as $\sqrt{s}$ and $n$ become $\gg 1$.

The $n$-Higgs amplitudes on the multi-particle mass thresholds read ({\it cf.} \eqref{eq:amplnh}),
\begin{eqnarray}
{\cal A}^{k\, \rm thr.}_{gg\to n\times h} &=& C_k \,\frac{\alpha_s}{\pi}\,(p_{1\mu}\epsilon_{1\nu}^a-p_{1\nu}\epsilon_{1\mu}^a)
(p_{2}^{\mu}\epsilon_{2}^{a\nu}-p_{2}^{\nu}\epsilon_{2}^{a\mu})\,
\left.\left(\frac{\partial}{\partial z}\right)^n \left(\frac{ y_t}{\sqrt{s}}\,\frac{z}{1-\frac{z}{2v}} \right)^k \,\right|_{z=0}
\nonumber\\
&\sim&
C_k \,\frac{\alpha_s}{\pi}\,\,y_t^k \,\left(\frac{1}{1+\varepsilon}\right)^{k-2}\,\left(\frac{1}{nM_h}\right)^{k-2}
\left.\left(\frac{\partial}{\partial z}\right)^n \left(\frac{z}{1-\frac{z}{2v}} \right)^k \,\right|_{z=0}
\,,
\label{eq:Aggnh}
\end{eqnarray}
where in the final expression we used the substitutions $(p_1+p_2)^2=s$ and $\sqrt{s}=(1+\varepsilon)nM_h$.
Of course the true threshold amplitude is obtained from \eqref{eq:Aggnh} by setting $\varepsilon=0$.

We now note that since 
\[
\left.\left(\frac{\partial}{\partial z}\right)^n h_{\rm cl}[z] \,\right|_{z=0} \,=\,
\left.\left(\frac{\partial}{\partial z}\right)^n \left(\frac{z}{1-\frac{z}{2v}} \right) \,\right|_{z=0}\,=\,
 \, n! \, \left(\frac{1}{2v}\right)^{n-1}\,,
\]
the same operation applied to the $k$-th power of the classical solution will lead in the large-$n$ limit to
\[
\left.\left(\frac{\partial}{\partial z}\right)^n \left(\frac{z}{1-\frac{z}{2v}} \right)^k \,\right|_{z=0}\,\to\,
 \, n^{k-1} \,n! \, \left(\frac{1}{2v}\right)^{n-k-1}\,.
\]
In particular, it can be verified that for $k=2$ (Boxes) the expression valid for all values of $n$ is
\[
\left.\left(\frac{\partial}{\partial z}\right)^n \left(\frac{z}{1-\frac{z}{2v}} \right)^2 \,\right|_{z=0}\,=\,
 \, (n-1)\, n! \, \left(\frac{1}{2v}\right)^{n-3}\,,
\]
and for $k=4$ (Hexagons) one gets,
\[
\left.\left(\frac{\partial}{\partial z}\right)^n \left(\frac{z}{1-\frac{z}{2v}} \right)^4 \,\right|_{z=0}\,=\,
 \, \frac{1}{6} (n^3-6n^2+11 n-6)\, n! \, \left(\frac{1}{2v}\right)^{n-5}\,,
\]

To summarise, the threshold amplitudes ($\epsilon=0$) in the large-$n$ limit read,
\[
\left.{\cal A}^{k\,\rm thr.}_{gg\to n\times h}\right|_{\epsilon=0}
 \,\to\, \,n\, n!\,  \left(\frac{\lambda}{2M_h^2}\right)^{\frac{n-1}{2}}\, \frac{C_k}{\kappa_k} \,\frac{\alpha_s}{\pi}\,\, M_h^2\,\left(\frac{2m_t}{M_h}\right)^k
\,,
\label{eq:Aggnh_thr}
\]
where $\kappa_2=1$ and $\kappa_4=6$. Above the threshold we should also include the multiplicative factor
$1/(1+\varepsilon)^{k-2}$ present on the right hand side of \eqref{eq:Aggnh}. Hence one can write for the 
above-the-threshold amplitude,
\[
{\cal A}^k_{gg\to n\times h} \,=\, \,n\,\left(\frac{1}{1+\varepsilon}\right)^{k-2}\, \frac{C_k}{\kappa_k} \,\frac{\alpha_s}{\pi}\,\, M_h^2\,\left(\frac{2m_t}{M_h}\right)^k\,\, {\cal A}_{h^*\to n\times h}
\,,
\label{eq:Aggnh_sch}
\]
and in addition, one should remember
that the tree-level amplitude on the right hand side will itself
contain dependence on the kinematics. For example, in the double-scaling large-$n$,
$\varepsilon \ll 1$ limit with $n\varepsilon$ held fixed,
the tree-level amplitudes 
in the Higgs model were computed in \cite{VVK2},
\[
{\cal A}_{h^*\to n\times h}\,=\, n!\,  \left(\frac{\lambda}{2M_h^2}\right)^{\frac{n-1}{2}}\, \exp\left[-\frac{7}{6}\,n\, \varepsilon\right]\,.
\label{ASSB}
\]
We will postpone the discussion of the full kinematic dependence for these processes to the next section.

\medskip

The main conclusions we would like to draw from the discussion up to now is that in the high-energy, large-$n$ limit
the dominant contributions to the multi-particle amplitudes are succinctly characterised by the set of EFT
vertices \eqref{eq:efteven} or generating functionals \eqref{eq:Aggnh} 
with even values of $k\ge 2$.  Further simplification occur in the highly-relativistic kinematics where $\varepsilon$ is large.
In this case, the factor $1/(1+\varepsilon)^{k-2} \ll 1$ in \eqref{eq:Aggnh_sch}
suppresses all contributions from $k>2$ -- hence in this case the dominant 
contributions come from the Boxes.

In the kinematic regime where $\varepsilon \lesssim 1$ all even-$k$ polygons contribute and are described by the 
amplitudes \eqref{eq:Aggnh_thr}. The constants $C_2$ and $C_4$ were computed in Table~\ref{tab:Cn} together with
$\kappa_2=1$ and $\kappa_4=6$. Hence the numerical prefactors for the boxes and the hexagons are fully accounted for.
But the main point of our analysis is that all even polygons contribute to the same $n$-dependence of the amplitudes in
 \eqref{eq:Aggnh_thr}, and the cross-sections at large-$n$ will have the characteristic same exponential behaviour 
 which will be determined in the following section.

\medskip
\section{Exponential form of the multi-particle cross-section}\label{sec:An}

Let us consider the multi-particle limit $n\gg k \approx 1$ and scale the energy 
$\sqrt{s} = E$ linearly with $n$,  $E \propto n$, 
keeping the coupling constant small at the same time, $\lambda \propto 1/n.$ 
Based on the characteristic form $\sim n!  \lambda^{n/2}$ of the multi-particle scattering amplitudes on and above the multi-particle thresholds, it was first pointed out in \cite{LRST}, and then argued for extensively in the literature,
that in this double-scaling limit the production cross-sections $\sigma_n$ have a characteristic exponential form,
\[
\sigma_n \,\sim\, e^{ \,n F(\lambda n, \,\varepsilon)} \,,
\quad {\rm for}\,\, n \to \infty\,,\,\, \lambda n ={\rm fixed}\,,\,\, \varepsilon ={\rm fixed}\,,
\label{eq:hg}
\]
where $\varepsilon$ is the average kinetic energy per particle per mass in the final state \eqref{eq:eps_n},
and $F(\lambda n, \varepsilon)$ is a certain a priori unknown function of two arguments, often referred to 
as the `holy grail' function for the multi-particle production. At tree-level, the dependence on $\lambda n$ and $\varepsilon$,
factorises into individual functions of each argument,
\[
F^{\rm tree}(\lambda n, \,\varepsilon) \,=\, f_0(\lambda n)\,+\, f(\varepsilon)\,,
\label{eq:nFtree}
\]
and the two independent functions are given by the following expressions in the Higgs model \eqref{eq:LSSB} 
(See Refs.~\cite{LRST,VVK2}):
\begin{eqnarray}
\label{f0SSB}
f_0(\lambda n)&=&  \log\left(\frac{\lambda n}{4}\right) -1\,, 
\\
\label{feSSB}
f(\varepsilon)|_{\varepsilon\to 0}&\to& f(\varepsilon)_{\rm asympt}\,=\, 
\frac{3}{2}\left(\log\left(\frac{\varepsilon}{3\pi}\right) +1\right) -\frac{25}{12}\,\varepsilon\,.
\end{eqnarray}
These formulae is the result of integrating the tree-level amplitudes expressions \eqref{ASSB}
over the Lorentz-invariant phase-space, 
$\sigma_n =\frac{1}{n!} \int \Phi_n  \, \left|{\cal A}_{n}\right|^2$, in the large-$n$ non-relativistic approximation.
In particular, the ubiquitous factorial growth of the large-$n$ amplitudes \eqref{ASSB} translates into the 
 $\frac{1}{n!} |{\cal A}_{n}|^2 \sim n! \lambda^n \sim e^{n\log(\lambda n)}$ factor in the cross-section, which determines the 
 function $f_0(\lambda n)$ in  \eqref{f0SSB}. The energy-dependence of the cross-section is dictated by
 $f(\varepsilon)$ in Eq.~\eqref{eq:hg}, and this function arises from integrating the $\varepsilon$-dependent factors in \eqref{ASSB}
 over the phase-space, giving rise to the small-$\varepsilon$ asymptotics in \eqref{feSSB}.
 While the function $f_0(\lambda n)$ is fully determined at tree-level, the second function, $f(\varepsilon)$, characterising the 
 energy-dependence of the final state, is determined by \eqref{feSSB} only at small $\varepsilon$, i.e. near the multi-particle threshold.
 
\medskip

\begin{figure}[t]
\begin{center}
\includegraphics[width=0.52\textwidth]{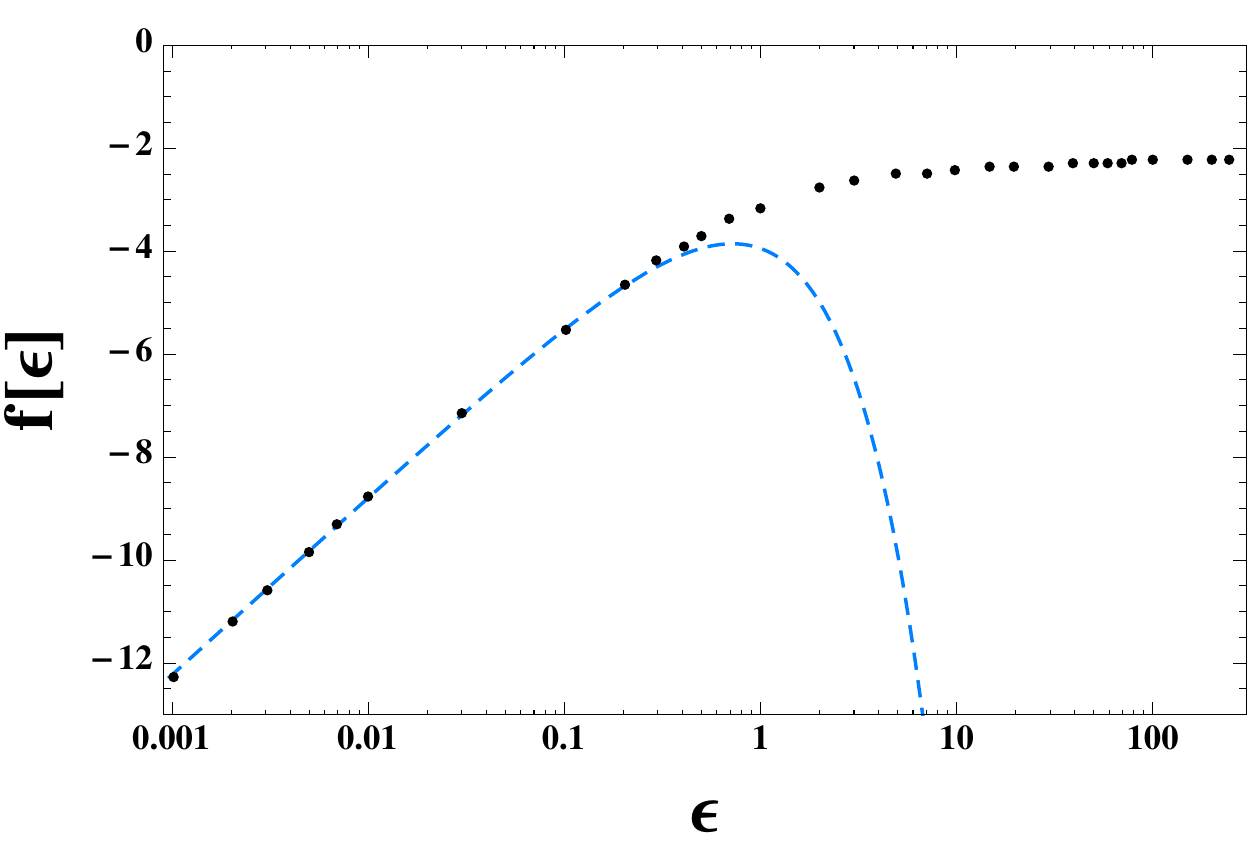}
\end{center}
\vskip-.6cm
\caption{
Plot of $f(\varepsilon)$ extracted from the $\log \sigma_7^{\rm tree} /\sigma_6^{\rm tree}$ MadGraph data in Ref.~\cite{VVK3}.
The results perfectly match $f(\varepsilon)_{\rm asympt}$
for $\varepsilon < 1$ shown as the dashed curve.
In the UV regime 
the function asymptotes to a constant
$f(\varepsilon=250) \simeq -2.2$
}
\label{fig:feps}
\end{figure}

The function $f(\varepsilon)$ in the {\it entire range} of $0\le \varepsilon < \infty$ was obtained in Ref.~\cite{VVK3} 
from the direct computation of tree-level perturbative cross-sections with 
up to $n=7$ Higgs particles, combined with the known large-$n$ scaling of the cross-section as defined
by $f_0(\lambda n)$ in Eqs.~\eqref{eq:nFtree}-\eqref{f0SSB}. 
The function $f_0(\varepsilon)$ is shown in Fig.~\ref{fig:feps}.
This plot also shows a perfect match to the known $f(\varepsilon)_{\rm asympt}$
expression \eqref{feSSB} at $\varepsilon < 1$, which is shown as a dashed curve in light blue.

\medskip

Having determined the $n$-independent kinetic energy function $f(\varepsilon)$ allows to us to compute multi-particle 
cross-sections at any $n$ in the large-$n$ limit. 

Let us now consider the effect of higher loop corrections in a single tree process\footnote{Therefore 
including only the factorable loop for process with $k>1$.}
 $h^\star\to n \times h$.
 It was shown in Ref.~\cite{LRST}, based on the analysis of leading singularities of the multi-loop expansion around singular generating functions in scalar field theory,
that the 1-loop correction exponentiates and results in the modified expression for $f_0$ 
\[
f_0(\lambda n)^{\rm NLO-resummed} \,=\, \log\left(\frac{\lambda n}{4}\right) -1 \,+\, \sqrt{3}\, \frac {\lambda n}{4\pi}\,.
\label{eqnl}
\]

Finally we can now use the expression for the EFT vertex \eqref{eq:Aggnh_thr}
and represent the cross-section via
\[
\sigma_n \,=\,  K_k\, \frac{n^2}{s}\,
\left(\frac{1}{1+\varepsilon}\right)^{2(k-2)}
\,e^{n\left(f_0(\lambda n)\,+\, f(\varepsilon)\right)}\, =\, K_k\
\frac{1}{M_h^2}\,
\left(\frac{1}{1+\varepsilon}\right)^{2k-2}
\,e^{n\left(f_0(\lambda n)\,+\, f(\varepsilon)\right)}
\,.
\label{eq:sigmaff}
\]
In the above formula, all even values of $k\ge 2$ can contribute. 
We think of $k$ as a characteristic $k$-value for which the prefactor, denoted as $K_k$ in \eqref{eq:sigmaff},
would be maximal.
This prefactor contains numerical factors appearing in the squared amplitude \eqref{eq:Aggnh_thr},
\[
K_k \sim\, \left(\frac{C_k \alpha_s}{\kappa_k \pi} \right)^2\,\left(\frac{2\sqrt2m_t}{M_h}\right)^{2k}
\,\simeq\,
\begin{cases}
\quad 0.1 & :\,\,k=2\\
\quad 20 & :\,\,k=4\, .
\end{cases}
\]
%
 \begin{figure}[h!]
\qquad\qquad\qquad\qquad\quad\includegraphics[width=0.6\textwidth]{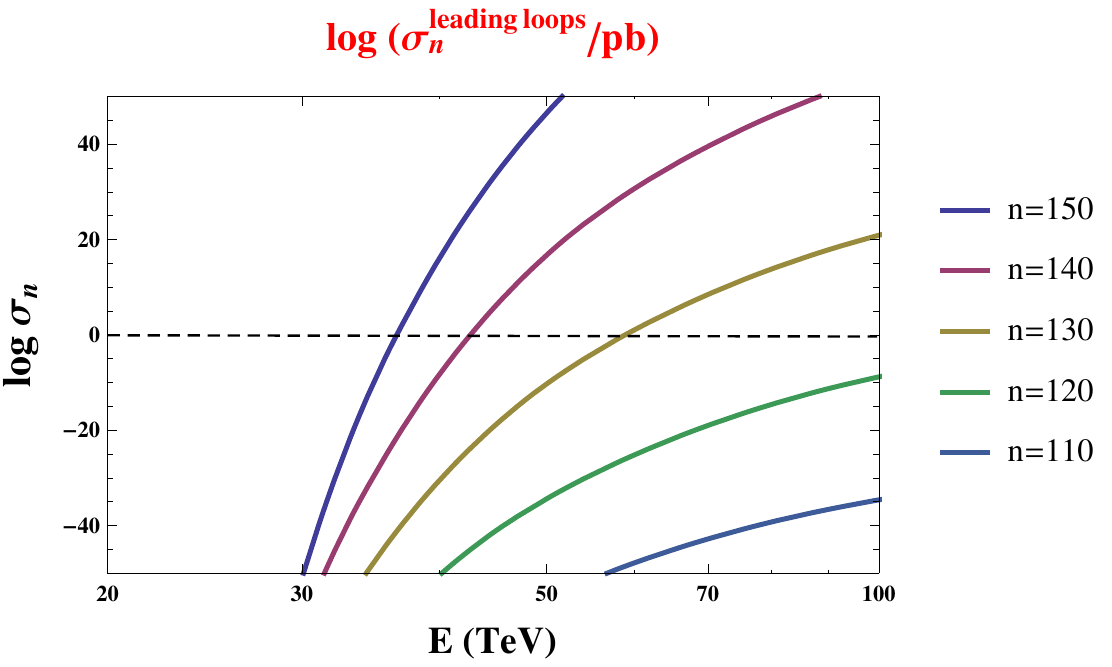}
\caption{
The logarithm of the cross-section \eqref{eq:sigmaff2} (in picobarns) 
is plotted as the function of energy for a range of final-state multiplicities between $n=110$ and $n=150$. 
We used the 1-loop-improved expression \eqref{eqnl} for $f_0(\lambda n)$. The plot corresponds to $k=2$ (boxes)
but there are only slight visible differences with the higher case $k=4$ (hexagons) {\it cf.} Fig.~\ref{fig:f7}.
}
\label{fig:f6}
\end{figure}
%
For a practical calculation we can take $k=2$ and $k=4$ and plot the logarithm of the cross-section 
\eqref{eq:sigmaff} for these two cases using the formulae,
\[
\log(\sigma_n/{\rm pb})
\,\simeq\,
\begin{cases}
 n\left(f_0(\lambda n)\,+\, f(\varepsilon)\right) -2\log(1+\varepsilon)+8 & :\,\,k=2\\
n\left(f_0(\lambda n)\,+\, f(\varepsilon)\right) -6\log(1+\varepsilon)+13 & :\,\,k=4\, .
\end{cases}
\label{eq:sigmaff2}
\]
The plot in Fig.~\ref{fig:f6} plots these cross-sections as the function of energy for a 
range of final-state Higgs multiplicities between $n=110$ and $n=150$. The  
specific form of the prefactor for $k=2$ and $k=4$ has an almost negligible effect on the logarithm of the
cross-sections, and the plot depicts only the minimal $k=2$ case.
%
 \begin{figure}[h!]
\qquad\qquad\qquad\qquad \qquad \includegraphics[width=0.75\textwidth]{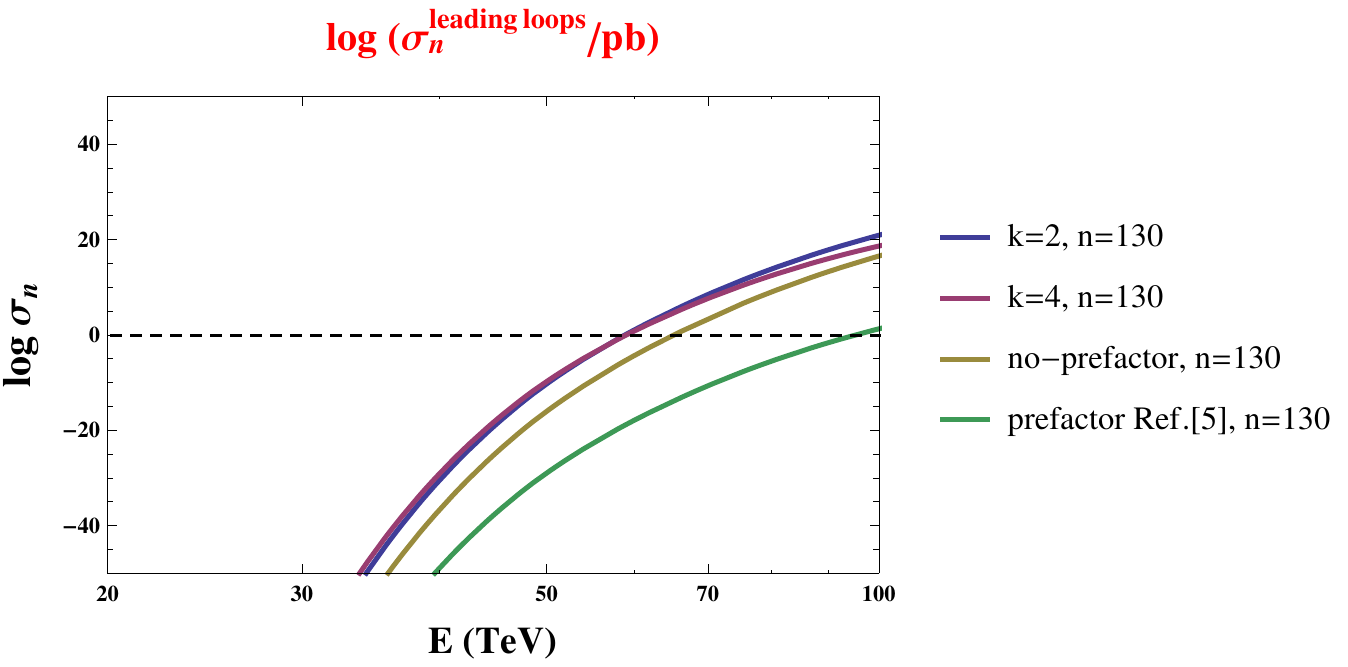}
\caption{
The logarithm of the $n$-particle cross-section \eqref{eq:sigmaff2} for $n=130$ shown as the function
of energy. The four contours represent four different choices for the prefactor: the first two correspond to the $k=2$ and $k=4$
expressions in \eqref{eq:sigmaff2}, the third contour contains no prefactor, while the fourth case depicts the triangle EFT formfactor
used in Eqs.~(3.12)-(3.13) of Ref.~\cite{VVK3}.
}
\label{fig:f7}
\end{figure}
%
In Fig.~\ref{fig:f7} we choose a `relatively low' final state Higgs multiplicity of $n=130$ and show the 
limits derived in the earlier work \cite{VVK3} (based on using the dimension-5 EFT vertices with form-factors)\footnote{The choice 
$n=130$ is motivated by being on the very edge
of potential observability $\log \sigma_n \to 1$ at 100 TeV in the set-up of \cite{VVK3}.}
versus the cross-section expressions obtained with prefactors based on boxes ($k=2$), hexagons ($k=4$), and with no
prefactors. We conclude that there is little difference in practice between the latter three cases, while they all show 
improvement relative to the dimension-5 EFT vertices. 

The fact that the leading-loop correction to the tree-level amplitudes on multi-particle mass-thresholds exponentiate
and result in Eq.~\eqref{eqnl} is a well-established fact \cite{LRST,LST,Libanov:1996vq} based on a complete multi-loop computation in the background of
the classical solution \eqref{sol-SSB} at its singular point. The fact that loop correction (the last term in \eqref{eqnl})
is positive in the Higgs theory \eqref{eq:LSSB} is instrumental in lowering the energy scale where the cross-sections 
stop being small to ${\cal O}(100\, {\rm TeV})$ and $n \sim 130$. If one wished to use the pure tree-level expression
for $f_0(\lambda n)$ in Eq.~\eqref{f0SSB}, both the desired energy scale and the multiplicity will increase by an order of magnitude,
as can be seen from Figure 5 of Ref.~\cite{VVK3}.

Of course, one should keep in mind that there are even higher-order corrections to the exponent of the 
multi-particle cross-sections arising from the higher loop effects. Moreover, only the loop inside the trees are included and not those connecting different trees.  Hence the use of the 1-loop-improved expression
in  \eqref{eqnl} should be seen as an optimistic
phenomenological model. In general 
the higher-order effects of loop exponentiation will amount to
\[
f_0(\lambda n)^{\rm all \,loops} \,=\,  \log\left(\frac{\lambda n}{4}\right) -1 \,+\, \sqrt{3}\, \frac {\lambda n}{4\pi}
 \,+\, {\rm const} \left( \frac {\lambda n}{4\pi}\right)^2  \,+\, {\rm const'} \left( \frac {\lambda n}{4\pi}\right)^3 \,+\, \ldots
\,, 
\]
and can change the cross-sections contours in Fig.~\ref{fig:f6}.\footnote{Note that the loop expansion parameter
$\frac{\lambda n}{4\pi}$ is $\simeq 1$ for $n=100$ and $\simeq 1.4$ for $n=140$.}
Furthermore, the exponentiation of loop-level effects which was proven for amplitudes on mass thresholds, is not the full story; 
one expects that there are additional multi-loop contributions to the holy grail function $F(\lambda n,\varepsilon)$ 
which depend on both $\lambda n$ and $\varepsilon$ and cannot be separated into $f_0(\lambda n)$ and $f(\varepsilon)$.

However what we can state with certainty is that the perturbation theory becomes strongly-coupled and
 breaks down for multi-particle processes when of the order of 130 
higgses are produced at energies $\sim {\cal O}(100 \, {\rm TeV})$. 

We also expect that a similar conclusion will hold for the production of the massive vector bosons ($W$'s and $Z$'s) in the 
electroweak gauge sector. In the preliminary studies in Ref.~\cite{VVK1} it was shown that similarly to the case of massive scalars,
the high-multiplicity production of the longitudinal components of the massive vector bosons also exhibits the factorial growth  
of the tree-level amplitudes. We plan to return to these studies in near future.

\section{Convolution with Parton Distribution Functions}
\label{sec:PDFs}

\begin{figure*}[h!]
\begin{center}
\includegraphics[width=0.52\textwidth]{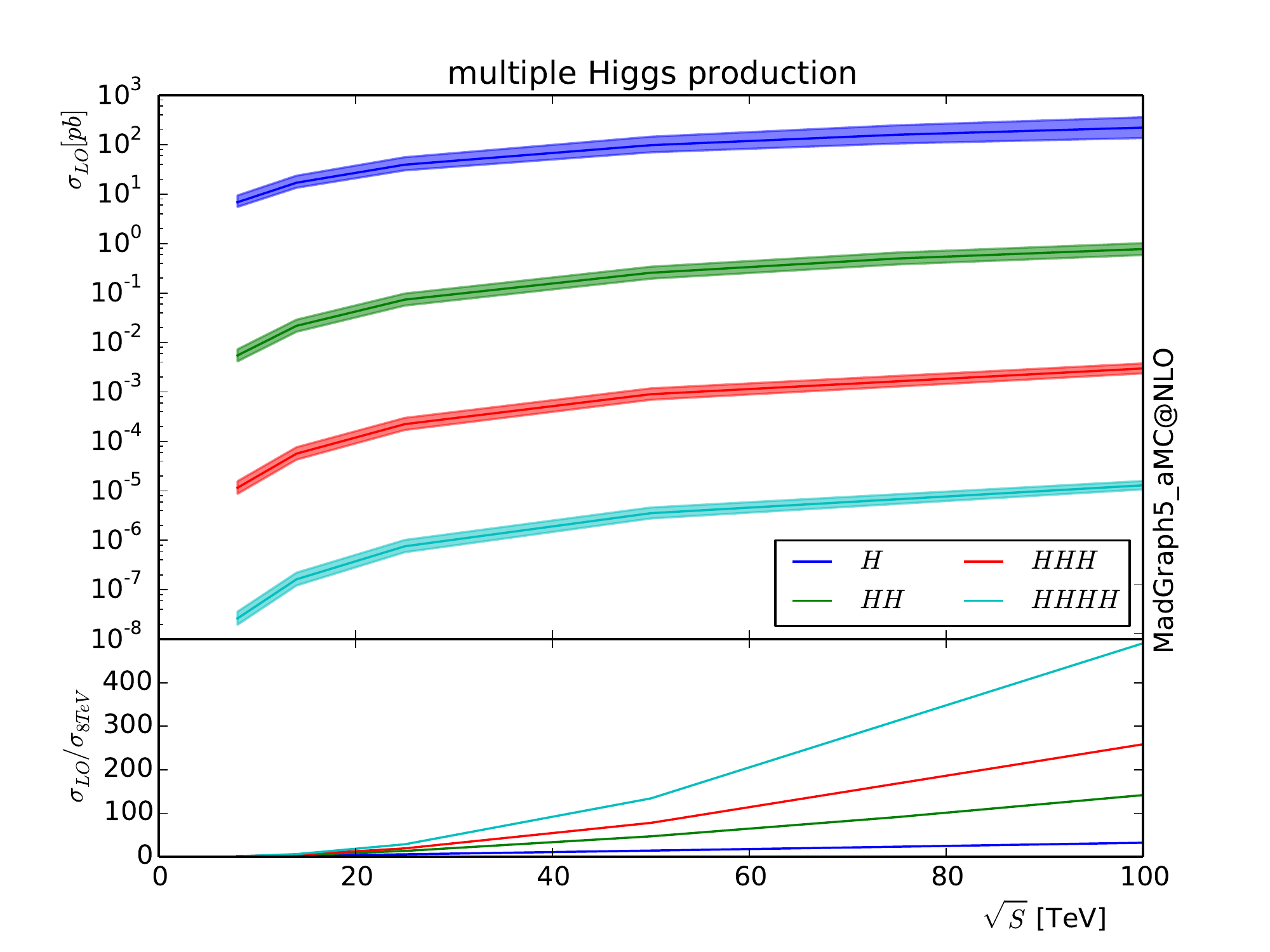}
\end{center}
\vskip-.4cm
\caption{Leading Order cross-sections with up to 4 Higgs bosons computed by Madgraph5\_aMC@NLO using MSTW2008 PDF \cite{Martin:2009iq}. The band correspond to the scale systematics by changing the scales by factor of two.}
\label{fig:pdf_4h}
\end{figure*}

The PDFs have a huge influence for the production of a few Higgs bosons as can be seen in Fig.~\ref{fig:pdf_4h} where we plot the 
Leading Order cross-sections with up to 4 Higgs bosons computed by
Madgraph5\_aMC@NLO. The lower panel in this figure shows the ratio of these cross-sections to the ones obtained at 8 TeV. 
The larger the number of Higgses produced, the bigger is the enhancement with the collider energy, 
as expected from the PDF enhancement effect of a more energetic collider.  On the other hand, the cross-sections drop by a few orders
of magnitude for each extra Higgs in the final state. As a matter of fact, the PDF rapid fall heavily suppresses the rate of processes 
with a higher threshold.

\begin{figure*}[h!]
\begin{center}
\includegraphics[width=0.49\textwidth]{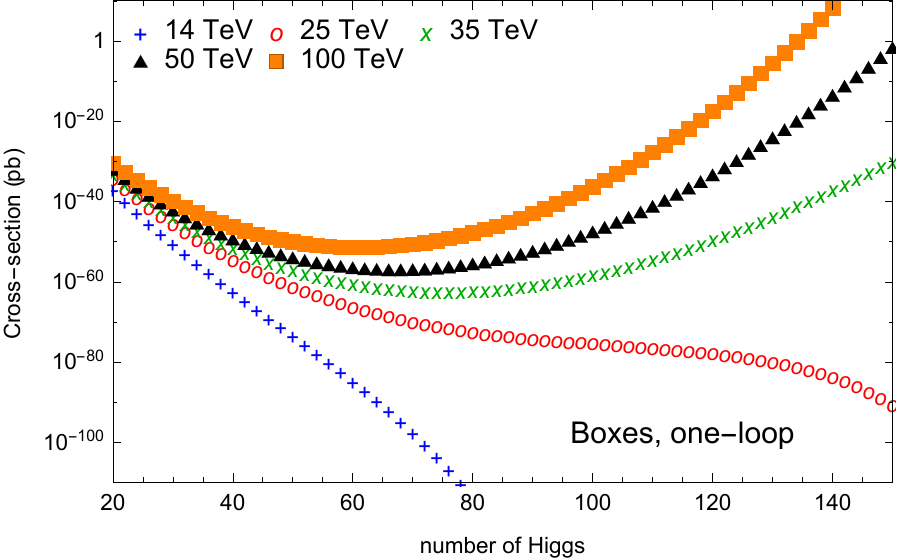}
\includegraphics[width=0.49\textwidth]{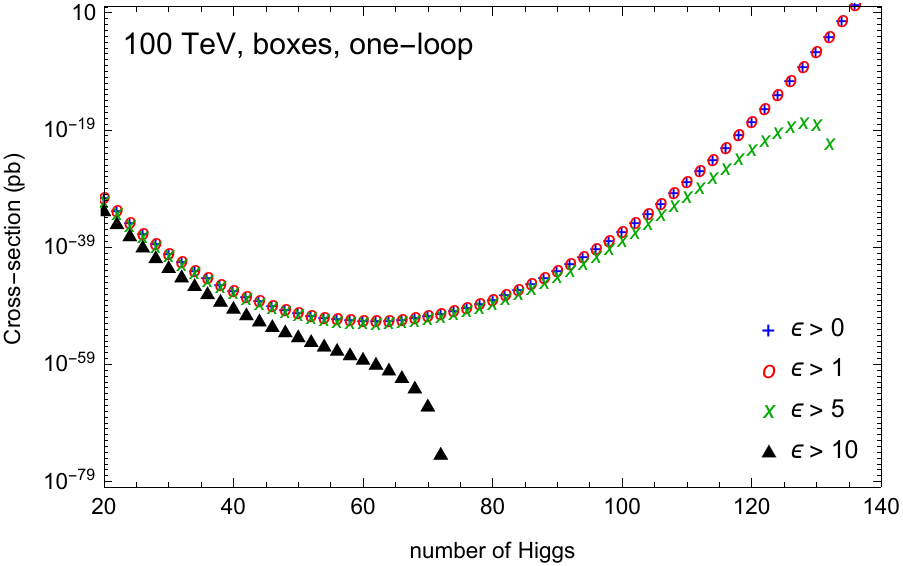}
\end{center}
\caption{Left panel: Cross-sections for multi-Higgs production \eqref{eq:sigmaff} at proton colliders including the PDFs for 
different energies of the proton-proton collisions plotted as the function of the Higgs multiplicity. Only the contributions from the boxes
are included. The right panel illustrates the dependence on average energy variable $\varepsilon$
by applying a sequence of cuts on $\varepsilon$ at 100 TeV.}
\label{fig:pdf}
\end{figure*}

Although the energy of the exponential growth for the production of many Higgses is within the reach of the FCC, one could wonder if this effect is not completely washed away by PDF suppression due to the very high  threshold. 
We show in Fig.~\ref{fig:pdf} that this is not the case, and the rapid growth of partonic rates leads to picobarn cross-sections already 
for a 50 TeV collider for the production of $\gtrsim$140 bosons. For lower energy collider, the PDFs are killing the cross-section before reaching the fast growth regime. On the right plot of  Fig.~\ref{fig:pdf}, we display the cross-sections with a lower cut on the average kinetic energy $\varepsilon$ per particle per mass. Since this variable is directly related to the partonic centre of mass energy,
\begin{equation}
\sqrt {\hat s}  = (\varepsilon+1) n M_h,
\end{equation} 
this cut is equivalent to a cut on the partonic energy of the collision. It should be noted that the largest contribution to the cross-section occurs when $\varepsilon$ is $\sim$ a few (neither large nor small). Much higher value of $\varepsilon$ are just not kinematically available. On the other side, the threshold is quite suppressed such that the contribution of the region $\varepsilon\lesssim1$  is also negligible.
The plot  only includes the contribution of the boxes since these are expected to be dominant for large values of $\varepsilon$, 
and are of the same size as the other even polygons for $\varepsilon \sim 1$, as shown in the previous section.

 \section{Conclusions}
 \label{sec:conc}
 
We have carried out a detailed study of multi-Higgs production processes in the gluon fusion channel 
in the high energy regime relevant to Future Circular hadron colliders and in the high-Higgs-multiplicity limit.
Our results are based on the computation of the leading polygons -- the triangles, boxes, pentagons and hexagons --
to the scattering processes, further combined with the subsequent branchings to reach high final state multiplicities.

We find that the characteristic energy and multiplicity scales where these perturbative rates
become observable and grow exponentially with increasing energy are within the  50 and 100 TeV regime with of order of 130 Higgses (or more) in the final state.
This is the regime where a dramatic change away from the usual weakly-coupled perturbative description of the electro-weak
physics should occur. One can speculate that this is related to transitioning to 
a classicalization regime \cite{Dvali:2010jz,Dvali:2013vxa} (albeit in non-gravitational QFT settings)
where the dominant processes above the critical energy scale correspond to the higher and higher numbers of the 
relatively soft Higgs and vector bosons appearing in the final state (before their decay). It is not expected that the perturbation theory 
would be a valid description in this regime, but it does provide an indication for the critical values of the energy and occupation numbers.

\section*{Acknowledgements}

We would like to thank Gia Dvali, Cezar Gomez, Valentin Hirschi and Joerg Jaeckel for enlightening discussions. 
This work is supported by STFC through the IPPP grant.
OM and CD are Durham International Junior Research Fellows.
Research of VVK is supported in part by a Royal Society Wolfson Research Merit Award. 

\newpage

\startappendix
\Appendix{Tree-level amplitudes at threshold}

Here we will provide a brief overview of the generating functions approach for computing
tree-level scattering amplitudes on multi-particle mass thresholds. This elegant formalism pioneered by Lowell Brown in Ref.~\cite{Brown} 
is based on solving classical equations of motion and bypasses the summation over individual Feynman diagrams. The overview below is included primarily for the reader's convenience, our presentation follows closely an earlier discussion of the Brown's technique in Section 2 of Ref.~\cite{VVK1}. 

The amplitude ${\cal A}_{1\to n}$ for a scalar field $\phi$ to produce $n$ particles with mass $M$ and momenta $p_1^\mu,\ldots p_n^\mu$,
is found by taking the matrix element of $\phi$ between the vacuum states in the presence of 
an external source, $\rho(x)$, $\langle 0_{\rm out} |\phi(0)| 0_{\rm in}\rangle_\rho$,  
differentiating it $n$ times with respect to the source $\rho$, and applying the LSZ reduction,
\[ \langle n|\phi(x)| 0\rangle \,= \lim_{\rho\to0}\, \left(
\prod_{j=1}^{n}\,\lim_{p_j^2\to M^2}\int d^4 x_j e^{ip_j \cdot x_j}(M^2-p_j^2) \frac{\delta}{\delta\rho(x_j)}\right)
\langle 0_{\rm out} |\phi(x)| 0_{\rm in}\rangle_\rho\,.
\label{eq:A1}
\]
The approach of course is general, but for concreteness we will consider first the simplest scalar $\phi^4$ field theory
with the Lagrangian (including the source term $\rho\, \phi$),
\[
{\cal L}_\rho(\phi) \,= \, \frac{1}{2} \left(\partial \phi\right)^2 - \frac{1}{2} M^2 \phi^2 - \frac{1}{4} \lambda \phi^4
 +\, \rho\, \phi\,,
\label{eq:A2}
\]
We now make use of two simplifying conditions which will reduce dramatically the technical complexity of the problem.
The first simplifying point is that we intend to sum up only the tree-level processes, hence we can work at the zeroth order
in the loop expansion parameter $\bar{h}$. This is captured by the classical dynamics. Specifically, the 
tree-level approximation is obtained by replacing the matrix element $\langle 0_{\rm out} |\phi(x)| 0_{\rm in}\rangle_\rho$ 
on the {\it r.h.s.}  of \eqref{eq:A1} by a solution 
$\phi_{\rm cl}(\rho;x)$ to the classical field equations.  corresponding to the Lagrangian ${\cal L}_\rho(\phi)$.
The presence of the source $\rho(x)$ in the Lagrangian ${\cal L}_\rho(\phi)$ implies that the classical field is a functional of  of the source 
and can be differentiated with respect to it, as required by \eqref{eq:A1}.

The second simplification arises from reducing the $1 \to n$ kinematics to the $n$-particle mass threshold limit. This corresponds
to making all outgoing particles to be produced at rest, $\vec{p}_j=0$. In this limit, it is sufficient to
consider the spatially-independent source $\rho(t)$. Specifically, before taking the $p_j^2\to M^2$ limit  in \eqref{eq:A1}, we set
all outgoing momenta to $p^{\mu}_j=(\omega, \vec{0})$, and choose $\rho(t)=\rho_0(\omega)\,e^{i\omega t}$. This amounts to the 
second substitution on the {\it r.h.s.}  of \eqref{eq:A1}:
\[
(M^2-p_j^2) \frac{\delta }{\delta\rho(x_j)} \, \phi_{\rm cl}(\rho;x) \, \rightarrow \, (M^2-\omega^2) \frac{\delta  }{\delta\rho(t_j)} \, \phi_{\rm cl}(\rho;t)
= \, \frac{\delta }{\delta z(t_j)} \, \phi_{\rm cl}(z;t)
\,. \label{eq:A3}
\]
In the right-most part of the above equation we have absorbed the factor of $M^2-\omega^2$ into the definition of the source  
by writing $\rho_0(\omega) = (M^2-\omega^2) \, z_0(\omega)$ and defining the rescaled source variable $z(t)=\, z_0\,\,e^{i\omega t} $.
Importantly, one can now take the required on-shell limit $\omega \to M$ simultaneously with
sending $\rho_0(\omega)$ to zero such that $z_0$ remains finite \cite{Brown}, 
\[
z(t)=\,  \left(\frac{\rho_0(\omega)}{M^2-\omega^2}\, \,e^{i\omega t}\right)_{\omega \to M} \, \to\,\,
z_0\,\,e^{i M t}
\,.
\]
The resulting classical field $\phi_{\rm cl} (z(t))$  expressed as the function of the rescaled source $z(t)$,
now solves the {\it homogeneous} classical equation since we arranged for the source term $\rho(t)$ to vanish in our
double-scaling on-shell limit $\omega\to M$, $\rho_0 \to 0$. 

It then follows from Eqs.~\eqref{eq:A1}, \eqref{eq:A3} that the tree-level amplitude ${\cal A}_{1\to n}$ at the $n$-particle threshold is  
obtained by a simple differentiation of $\phi_{\rm cl} (z(t))$,
\[
{\cal A}_{1\to n}\,=\, \langle n|\phi(0)| 0\rangle \,=\, \left.\left(\frac{\partial}{\partial z}\right)^n \phi_{\rm cl} \,\right|_{z=0}
\,. \label{{eq:A5}}
\]
The generating function $\phi_{\rm cl} (t)$ is a solution of the ordinary differential equation without source; in the theory 
\eqref{eq:A2} the equation is 
\[
d_t^2\phi + M^2 \phi +\lambda\phi^3 \,=\, 0\,.
\label{{eq:A6}}
\]
To give the generating function of amplitudes at multiparticle thresholds, the solution must contain only the positive frequency 
components of the form $e^{+ i n M t}$ where $n$ is the number of final state particles in the amplitude ${\cal A}_{1\to n}$. This follows immediately from \eqref{{eq:A5}}.
Thus, the solution we are after is given by the Taylor expansion in powers of the complex variable $z(t)$,
\[
\phi_{\rm cl} (t) \,=\,
\sum_{n=1}^{\infty} a_n\, z(t)^n
\label{eq:A7}
\]
In the limit where interactions are switched off, $\lambda=0$, the correctly normalised solution is $\phi_{\rm cl} = z(t)$ and this fixes the first coefficient $a_1=1$ on the {\it r.h.s.} of \eqref{eq:A7}.  
As the solution contains only positive frequency harmonics, it is a complex function of Minkowski time. This also fixes the initial conditions of the 
solution, $\phi_{\rm cl} (t) \to 0$ as Im$(t)\to \infty$. In Euclidean time the solution is real. 

The Taylor expansion coefficients $a_n$ in \eqref{eq:A7} determine the actual amplitudes via \eqref{{eq:A5}}
giving ${\cal A}_{1\to n}\,=\, n! \, a_n$. The classical generating function approach of \cite{Brown} amounts to 
finding the $\vec{x}$-independent solution of the Euler-Lagrange equations
as an analytic function of $z$ in the form \eqref{eq:A7}, and computing the amplitudes via \eqref{{eq:A5}}.

The classical generating function for the theory defined by \eqref{eq:A2} is surprisingly simple and can be written in closed form \cite{Brown},
\[\phi_{\rm cl} (t) \,=\, \frac{z(t)}{1-\frac{\lambda}{8M^2}\, z(t)^2}\,.
\label{noSSB}
\]
It is easily checked that the expression in \eqref{noSSB} solves the classical equation \eqref{{eq:A6}} and has the correct form, $\phi_{\rm cl} = z + \ldots$ as $z\to 0$.
The corresponding tree-level amplitudes on mass-thresholds in the theory \eqref{eq:A2} are then given by 
\[
{\cal A}_{1\to n}\,=\, 
\left.\left(\frac{\partial}{\partial z}\right)^n \phi_{\rm cl} \,\right|_{z=0}
\,=\, n!\, \left(\frac{\lambda}{8M^2}\right)^{\frac{n-1}{2}}
\,,
\label{eq:ampln2}
\]
and exhibit the factorial growth with the number of particles $n$ in the external state.

\medskip

 This general approach is also readily applied to the theory \eqref{eq:LSSB} with the non-vanishing VEV relevant to the 
 high-multiplicity Higgs production studied in the present paper. In this case the classical equation is 
 given by Eq.~\eqref{cleq-SSB}
and one searches for the particular solution in the form $h_{\rm cl} = v + z + {\cal O}(z^2)$, where the $z^0$ term is the VEV.
Instead of solving the second-order ordinary differential equation \eqref{cleq-SSB}, one can consider an equivalent problem 
which results from computing the first integral of motion of the associated to the theory \eqref{eq:LSSB} Euclidean problem.
In this case one considers the first integral of motion -- the energy $E$ --  in the Euclidean time. $E$ must be constant on the 
classical trajectory, and in the case at hand, $E=0$,
\[
E:= \int dt \left( \frac{1}{2} (d_\tau h)^2 - \frac{\lambda}{4}\left(h^2-v^2\right)^2\right) \,=\, 0\,,
\]
where $d_\tau h$ denotes the derivative of the field with respect to the imaginary time $\tau=it$.
This amounts to solving the first-order differential equation,
\[ d_\tau h \,=\, \sqrt{\lambda/2}\, (h^2-v^2)\,,
\label{eq:E0}
\]
or in Minkowski time:
\[ -i d_t h \,=\, \sqrt{\lambda/2}\, (h^2-v^2)\,.
\label{eq:E0M}
\]
The general solution of the first-order differential equation
easily found by the separation of variables, and, in particular, the 
imaginary-time equation \eqref{eq:E0} is solved by the hyperbolic tangent. It's analytic continuation to real time
is
\[
h_{\rm cl} (t) \,=\, v\,\frac{1+\frac{z(t)}{2v}}{1-\frac{z(t)}{2v}} \, , \quad {\rm where} \quad z=z_0 e^{i \sqrt{2 \lambda} v t}\,.
\label{sol-SSB2}
\]
This is precisely the Brown's solution \eqref{sol-SSB} used in the body of the paper, and it can also be checked by 
direct substitution that the expression \eqref{sol-SSB2} solves both the original classical  second-order
differential equation \eqref{cleq-SSB}, and the first-order equation \eqref{eq:E0M}. Taylor expanding \eqref{sol-SSB2}
in $z$ gives
\[
h_{\rm cl} (t) \,=\,  v\,+\, 2v\,\sum_{n=1}^{\infty} \left(\frac{z(t)}{2v}\right)^n
\, ,
\label{gen-funh2}
\]
which has the $z$ correct boundary conditions $h_{\rm cl} = v + z + {\cal O}(z^2)$ at $z\to 0$.

\bigskip

\bibliographystyle{h-physrev5}

\end{document}